\documentclass[a4,aps,twocolumn,showpacs,floatfix,longbibliography]{revtex4-1}
\usepackage{graphicx}
\usepackage{dcolumn}
\usepackage{color}
\usepackage{amsmath}
\usepackage{here}
\usepackage{systeme}
\usepackage{xcolor}

\begin{document} 
\title{Phase slips driven by acoustic waves  in Bose-Einstein condensates with ring topology}
	\author{Y. Kuriatnikov$^{1,2}$, A. Olashyn$^{1}$,  A.I. Yakimenko$^{1}$}

	\affiliation{ $^1$ Department of Physics, Taras Shevchenko National University of Kyiv, 64/13, Volodymyrska Street, Kyiv 01601, Ukraine}
	\affiliation{ $^2$ Atominstitut, TU Wien, Stadionallee 2, 1020 Vienna,
Austria}

\begin{abstract}
Rotational superradiance is one of the most fascinating phenomena in black-hole physics. Here, with the aim of probing quantum properties of superradiance in the lab, we investigate the interaction of the acoustic waves with quantum vortices in  Bose-Einstein condensates (BEC) in the framework of dissipative mean-field model. We find the conditions of the acoustic-induced quantum phase slips in condensates with ring topology and discuss the possibility of observing an acoustic analogue of quantum superradiance in ultracold atomic gases.   
\end{abstract}

\maketitle

\section{Introduction} 


Rotation of any macroscopic body with internal degrees of freedom could amplify incident radiation. It is a generic process with energy transfer from one medium to another. Superradiance is a process of the amplification of some modes of a scattering wave (amplitude increase with constant frequency) after the scattering on a rotating object. Such a phenomenon exists in different variations in quantum mechanics, astrophysics and relativity\cite{bekenstein_schiffer_1998}.

In relativity, superradiance is intimately connected to tidal acceleration. As superradiance is related to rotation energy transmission, it does not necessarily have to be only relativistic phenomenon and is predicted in Newtonian dynamics as well \cite{brito_cardoso_pani_20151, cardoso_2013}. Rotational superradiance in curved spacetime has been described by Yakov Zel’dovich \cite{zeldovich1, zeldovich2}.  Quantization of this process makes rotating objects spontaneously radiating. A very similar mechanism is the cause of the black hole evaporation, even in the absence of rotation \cite{unruh_1981, hawking_1974}.

Superradiance has been predicted theoretically and observed experimentally in many classical and quantum physical systems \cite{brito_cardoso_pani_20151,2017NatPh..13..833T, faccio_wright_2019,faccio_wright_2019,demirkaya_dereli_guven_2019, demirkaya_dereli_guven_2020, marino_ciszak_ortolan_2009}. 
As known, circulations in superfluids are quantized, and amplification cannot happen without decay of persistent current (reducing of topological charge) \cite{federici_cherubini_succi_tosi_2006}. Phase slips in toroidal condensates \cite{2013PhRvL.110b5302W, yakimenko_bidasyuk_weyrauch_kuriatnikov_vilchinskii_2015, ramanathan_wright_muniz_zelan_hill_lobb_helmerson_phillips_campbell_2011} are accompanied by vortex transfer to the condensate periphery. There vortex decays and the rotational energy of the superflow converts into acoustic waves. Thus, acoustically driven phase slips can be treated as quantum analogue of rotational superradiance.   Acoustic wave dynamics and interaction with a vortex are quite similar in the classic \cite{ford_smith_1999, howe_2003} and quantum \cite{tsubota_kobayashi_takeuchi_2013, meppelink_koller_straten_2009, ville_saint-jalm_cerf_aidelsburger_nascimbene_dalibard_beugnon_2018, ghazanfari_mustecaplioglu_2014} systems. However, it is worth to mention that for the strong perturbations shock waves appear instead of acoustic waves \cite{wang_kumar_jendrzejewski_wilson_edwards_eckel_campbell_clark_2015}. Acoustic wave interaction with  vortex in BEC is dramatically  affected by quantization of angular momentum.   Here we address the following question: how does quantum nature of the vortices in BEC influence the superradiance?  

A very recent experiment \cite{2017NatPh..13..833T} in the water tank demonstrated the amplification of surface waves after scattering on draining vortex. Amplification also depends on the angular velocity of the vortex  \cite{basak_majumdar_20031, basak_majumdar_20032}. In homogeneous infinite condensate quantum  superradiance is forbidden \cite{anacleto_brito_passos_2011}.
 Furthermore, in the fluid, the drain is not always required for a vortex for superradiant sound scattering; this effect may occur even when the fluid density drops to zero at the vortex core \cite{slatyer_savage_2005}.

\begin{figure}[tbp]
	\includegraphics[width = \columnwidth]{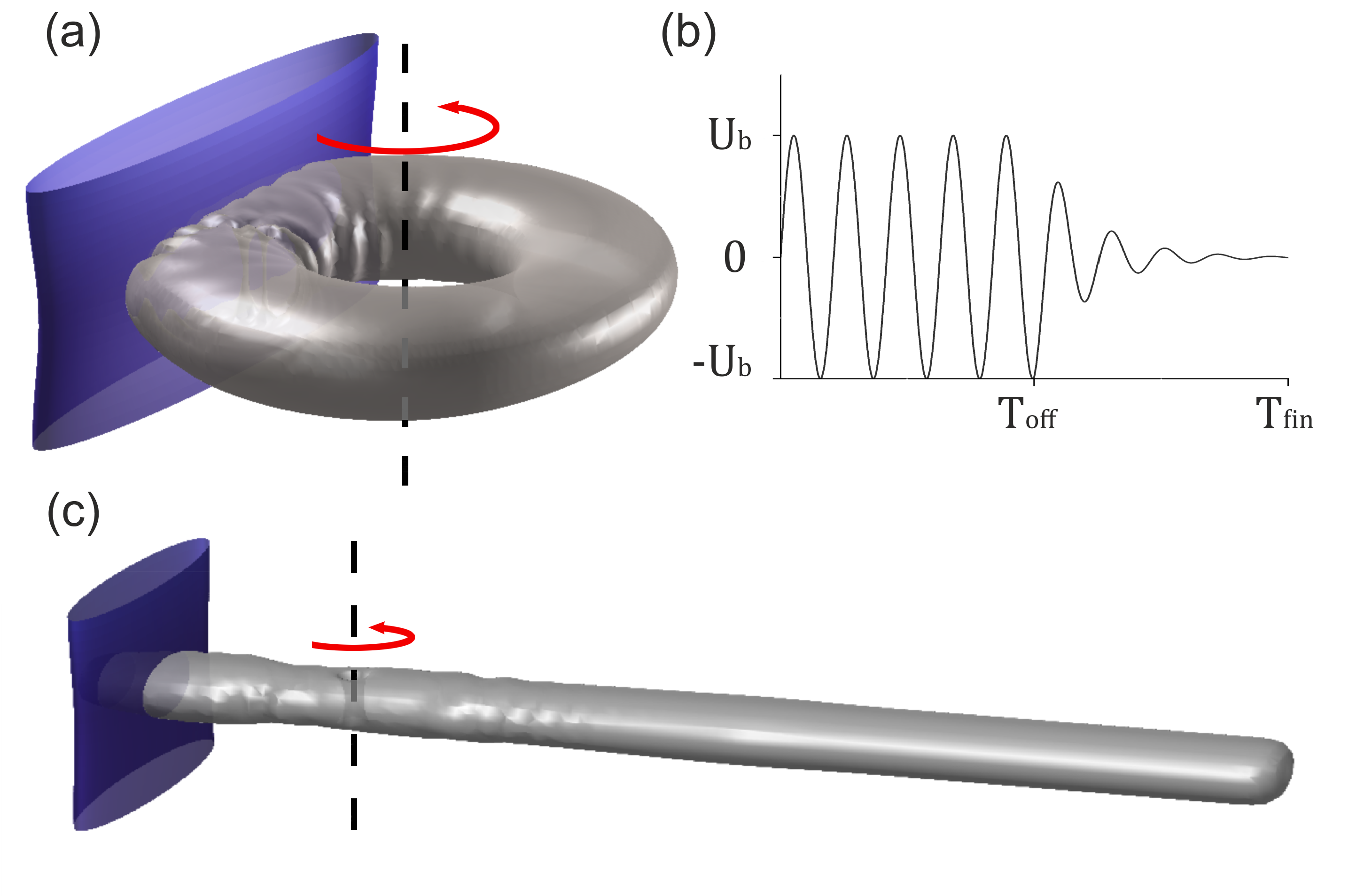}
	\caption{Schemes of the BECs used for investigation of the vortex-sound interaction. (a) Toroidal condensate perturbed by wide  Gauss-like potential. (b) Perturbation amplitude changing with time: oscillating until the moment T$_{off}$, then exponentially fast turns off and let the system relax before the final state measuring at the moment T$_{fin}$. (c) Elongated condensate tightly confined on $z$-direction, trapped in a box-like potential on $x$-direction, and parabolic potential on $y$-direction, with a mini-trap and the main axis for keeping a vortex thread. Black dashed lines represent the main axes of the single-charged imprinted vortex threads and red arrows indicate the flow direction. The blue walls are disturbing potentials that play the role of an acoustic wave generator. }
	\label{FIG:scheme}
\end{figure}

Here we investigate a quantum analogue of the experiment \cite{2017NatPh..13..833T} and suggest a setup for observation of quantum superradiance in ultracold atomic gases. We study the interaction of acoustic waves with a vortex trapped in a toroidal trap and quasi-2D and 3D geometry with a mini-trap on the main axis for holding vortex thread. 

Our paper is organized as follows. In Sec. \ref{section:model}, we define the model we use to investigate the system. In Sec. \ref{section:3d} we investigate dissipative Bose-Einstein condensate in a toroidal trap with additional external oscillating potential and with an imprinted vortex in purpose to explore the superradiance effect and its quantized properties. In Sec. \ref{section:2d}, neglecting the noise, we investigate the influence of the dissipation parameter $\gamma$ on damping decrement $\Gamma$  for acoustic waves extending along the 2D and 3D elongated condensates. On the main axis of the oblong condensate, with a Gauss-like potential, we create a trap smaller than a transverse width of the condensate with the imprinted vortex in it. We study a parametric area of acoustic wave's amplitude and frequency for observing vortex emission from the trap. We summarize our results in the concluding Sec. \ref{section:conc}.
\section{Model}
\label{section:model}

In the mean-field approximation, the dynamics of a system of weakly interacting degenerate atoms at finite temperature is describing by the dissipating Gross-Pitaevskii equation (dGPE):

\begin{equation}
\begin{aligned}
\label{EQN:GPE}
(i-\gamma)\hbar  \frac{\partial\psi(\textbf{r},t)}{\partial t} =\Big(&-\frac{\hbar^2}{2M}\Delta+V_{ext}(\mathbf{r},t) \\
& -\mu +  g|\psi(\textbf{r},t)|^2 \Big) \psi(\textbf{r},t).
\end{aligned}
\end{equation}

Here $\psi$ is a complex wave function of the condensate, $\mu$ is a chemical potential of the condensate, $\gamma$ is a phenomenological dissipation parameter and for physical systems $\gamma \ll 1$, for more complex models parameter $\gamma$ connected to the temperature of the condensate via fluctuation-dissipation theorem, $g=4\pi\hbar^2a_s/m$ is the coupling strength, m is the mass of the ${}^{23}Na$ atom, $a_s=2.75$ nm is the s-wave scattering length, $\hbar$ is reduced Planck constant. $V_{ext}(\mathbf{r},t)$ is an external potential contains of two parts: static harmonic trap $V_{trap}(\mathbf{r})$ and dynamic perturbation $V_{pert}(\mathbf{r},t)$. 

To find initial stationary ground state $\psi_{\textrm{GS}}$ we solve Eq. (\ref{EQN:GPE}) without perturbing potential ($V_{pert}(\mathbf{r},t)=0$) start with a random complex wave function $\psi(\mathbf{r},t)$ and use imaginary time propagation method. The essence of the method lies in the fact that evolution occurs in complex time (Wick rotation at the angle $\pi/2$). Imaginary time propagation essentially implements the steepest descent method in the energy space \cite{BAO2006836}.

To find the numerical solution of the deterministic GPE (\ref{EQN:GPE}) we use the split-step Fourier transform method \cite{agrawal_2013, 1937-5093_2013_1_1}. In all our simulations, the initial states have persistent current (single-charged vortex thread). To get this state, we imprint vortex into stationary solution without rotation using the following ansatz:
	\begin{equation}
	\label{EQN:vortices_anzatz}
\psi(x,y)=A\psi_{\textrm{GS}}(\textbf{r})\cdot\left\{\tanh\left\{\rho/\xi\right\}\right\}^{|m|}e^{im\theta},
\end{equation}
where $A$ is a normalization constant introduced to preserve the number of atoms,  $m$ is the topological charge of the imprinted vortex line ($m = +1$ for a single charged vortex and $m = -1$ for a single charged anti-vortex respectively), $(\rho, \theta)$ coordinates of the imprinted vortex in polar coordinate system, and $\xi$ is the healing length in the point of highest density of the condensate. We assume that the vortex line is parallel to the $z$-axis.

To improve the vortex form, we apply the imaginary time propagation method with a little number of iterations to all states with imprinted vortices, taking into account that the ITP decreases the state energy \cite{2006JCoPh.219..836B}. We have found that the coordinates of the vortex cores do not change significantly during the short imaginary time evolution, while the density shape near the vortices changes considerably \cite{2016PhRvA..94f3642S}.
 \begin{figure}[tbp]
	\includegraphics[width = \columnwidth]{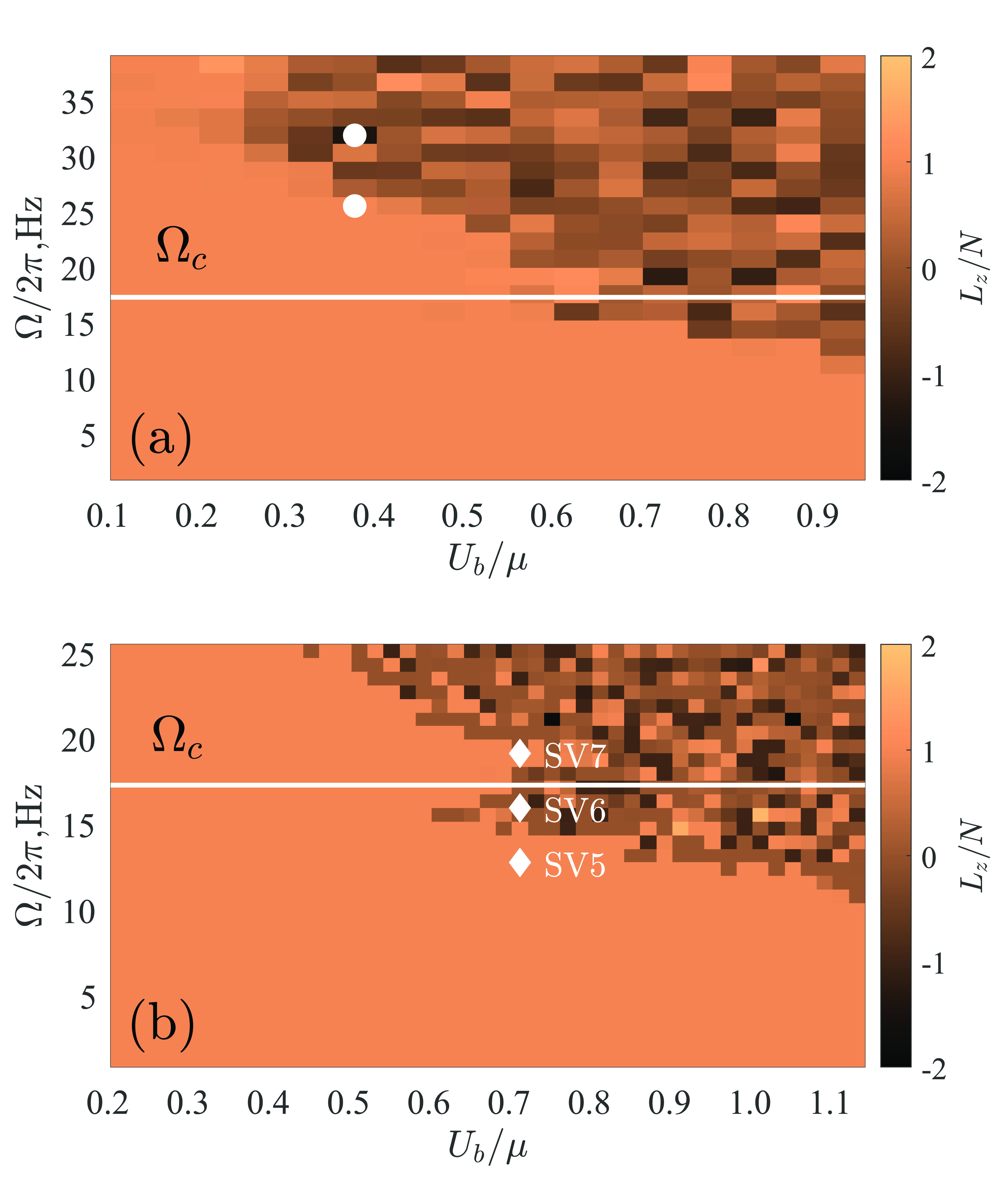}
	\caption{Final angular momentum per atom $L_z/N$ in torus for two different strength of dissipation \textbf{(a)} at the end of 2 sec evolution (perturbing potential is oscillating for 1.5 sec with frequency $\Omega$ and amplitude $U_b$ and then 0.5 sec relaxation without perturbation) of system with weak dissipation $\gamma=0.001$ and \textbf{(b)} at the end of 3 sec evolution (perturbing potential is oscillating for 2 and then relaxing during 1 sec) with dissipation $\gamma=0.0025$. Orange homogeneous area shows stable region where decay of single-charged persistent current under the influence of acoustic waves doesn't happen. Non-integer values of the angular momentum mean that at the time of the end of evolution, not all vortices/antivortices left the annulus. The key stages of evolution with perturbing parameters that correspond to white circles are shown in  Fig. \ref{FIG:evolution3d}. Full simulations for parameters SV5, SV6 and SV7 can be found in supplemental material videos.}
	\label{FIG:phaseslip3d}
\end{figure} 

\section{Wave propagation in toroidal 3D condensate}
\label{section:3d}

\begin{figure}[tbp]
	\includegraphics[width = \columnwidth]{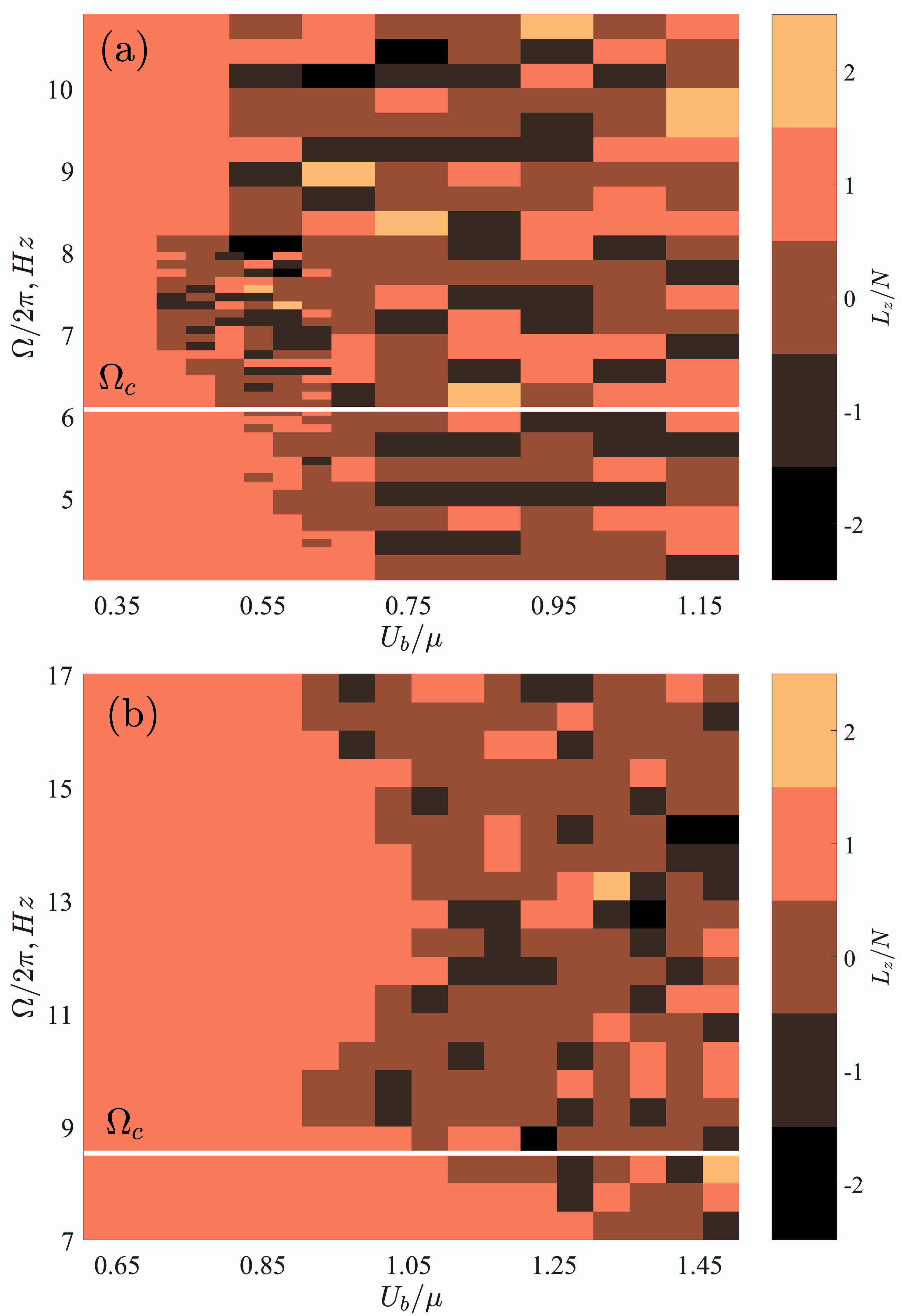}
	\caption{Final angular momentum per atom $L_z/N$ in torus at the end of 3 sec evolution (perturbing potential is oscillating for 2 and then relaxing during 1 sec) with dissipation $\gamma=0.0025$ For two different rings: (a) $N=6\cdot 10^5$, $(\omega_r,\omega_z)=2\pi\cdot(123,600)$ Hz, $R=47.12$ ${\mu}$m; (b) $N=2\cdot 10^5$, $(\omega_r,\omega_z)=2\pi\cdot(60,2400)$ Hz, $R=52.04$ ${\mu}$m. Orange homogeneous area shows stable region where decay of single-charged persistent current under the influence of acoustic waves doesn't happen, black dashed line shows critical frequency.}
	\label{FIG:bigring}
\end{figure}

\begin{figure*}[tbp]
	\includegraphics[width = \textwidth]{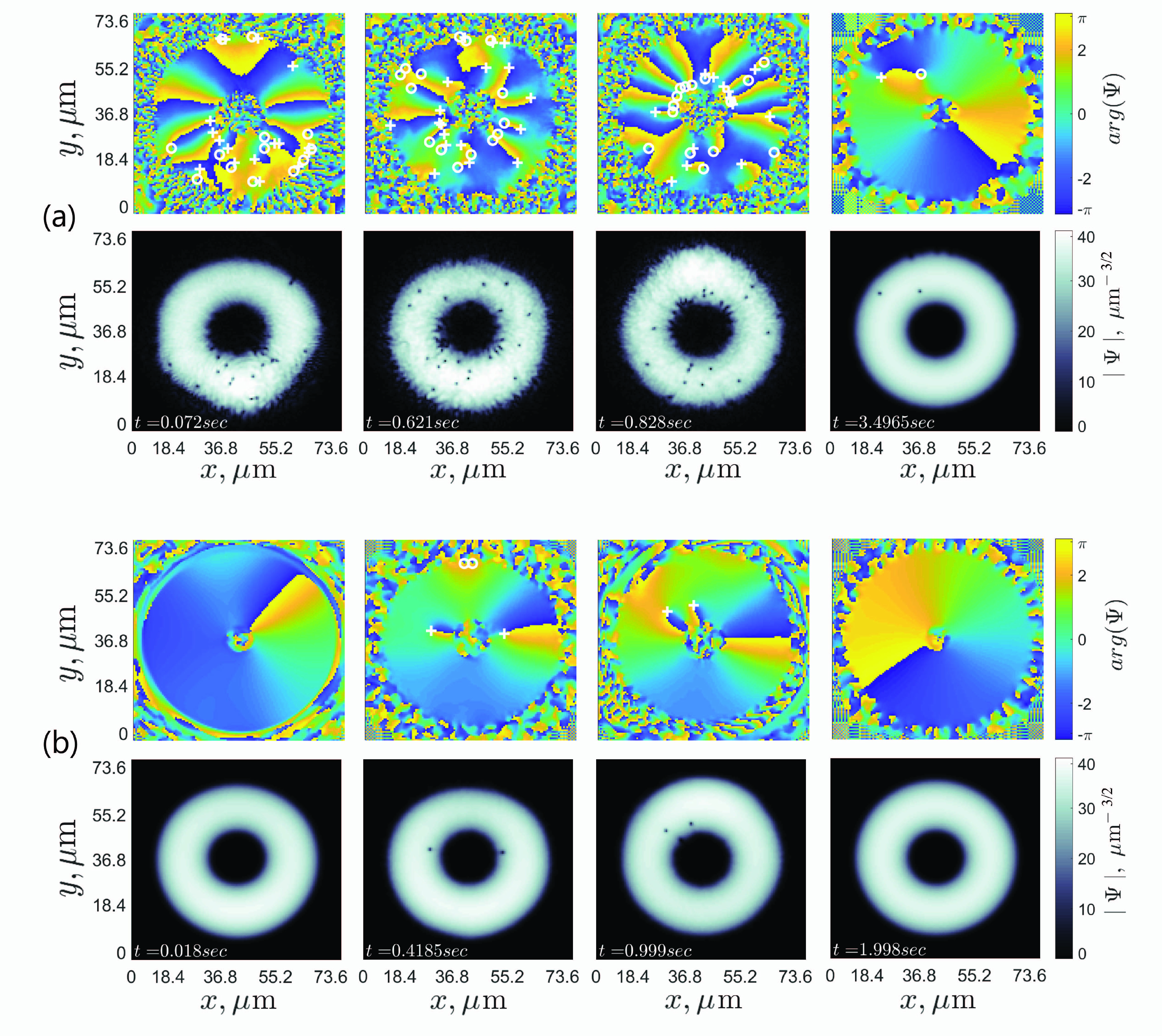}
	\caption{Snapshot series (a) and (b) show density and phase distribution evolution of the 3D toroidal condensate (in slice $z=0$) with periodic perturbation. Both series of shots correspond to the same amplitude of perturbing potential ($U_b/\mu=0.35$) but different frequencies that are marked as white circles in angular momentum per atom diagram (see Fig. \ref{FIG:phaseslip3d}). White crosses in phase profiles show the cores of the vortex threads, white circles - cores of antivortex threads. Time is shown in the left bottom corner of each density profile. (a) evolution in unstable area \textit{$\Omega=195$ }Hz: Perturbing potential excites large number of vortex-antivortex pairs on inner surface of torus; (b) evolution in stable area \textit{$\Omega=155$ }Hz: Perturbing potential may excite some vortices but they do not demonstrate complicated dynamics and do not destroy persistent current.}
	\label{FIG:evolution3d}
\end{figure*} 

 We consider toroidal BEC of $N=6\cdot 10^5$ atoms in the  harmonic trap with frequencies $(\omega_r,\omega_z)=2\pi\cdot(123,600)$ Hz, and peak-density ring radius $R=19.23$ ${\mu}$m; two-particle interaction parameter $g=0.018$,  and dissipation parameter $\gamma=0.001$.  We use these physical parameters because they are related to the experiment with the same geometry (see Ref. \cite{2013PhRvL.110b5302W}).

Dissipative system is non-conservative; therefore, chemical potential $\mu(t)$ is adjusted after each step to conserve number of particles $N$. The external potential consists of static harmonic trap and periodic quantum-mechanical "piston" perturbation See Fig. \ref{FIG:scheme} (a). Trapping potential is harmonic on $r$ and $z$ directions:  

	\begin{equation}
	\label{EQN:ext_pot_3d}
	V_{trap}(\mathbf{r})=\frac{M\omega_r^2(r-R)^2}{2}+\frac{M\omega_z^2 z^2}{2}.
	\end{equation}

The perturbation potential is periodic in time (turns off after the moment $t_{off}$) wide elliptical Gauss-like beam (See Fig. \ref{FIG:scheme} (b)) which main axis is tangent to the outer ring of the torus:

	\begin{equation}
	\label{EQN:perturbation_3d}
V_{per}(\mathbf{r},t)=\alpha U_{b} \exp\left\{{-\frac{x^2}{{w_x}^2}-\frac{(y-R_2)^2}{w_y^2}}\right\} \sin(\Omega t),
	\end{equation}
where $U_b$ is the amplitude of the perturbation, $r=\sqrt{x^2+y^2}$, $R_2$ is estimated outer Thomas-Fermi radius of the tor, parameters $w_x=2R_2$ and $w_y=R_2-R$ make perturbation thin in x-direction and wide in y-direction,

	\begin{equation}
	\label{EQN:alpha}
    \alpha(t)=\left\{
 \begin{array}{ll}
  1, t\leq t_{off}\\
 e^{-400(t-t_{off})}, t>t_{off}
\end{array}
\right.
	\end{equation}

In our simulations we use physically small parameter $\gamma = 0.001$ that provides relaxation of the system but does not remove short-wavelength excitation as can be seen in Fig. \ref{FIG:evolution3d} (a). Value of dissipation parameter $\gamma$ does not significantly effect the edge of phase-slip, as it can be seen in Fig. \ref{FIG:phaseslip3d} where map of after-evolution angular momentum per atom $L_z/N$ is shown in the same system with the same perturbation both with different dissipation parameter $\gamma = 0.001$ and $\gamma = 0.0025$. It mostly effects short-wavelength excitation and time of system's relaxation after turning off the "piston".

We have made numerous simulations for multi-charged systems. Such systems may be unstable and decay with time \cite{yakimenko_bidasyuk_weyrauch_kuriatnikov_vilchinskii_2015}. Presence of static repulsive potential destroys the axial symmetry and creates a weak-link in the energy barrier that separates states with different topological charge decreasing stability level comparing to the 'perfect' ring. In a static system with perturbing potential locked at its highest amplitude, the last stable rotating state has topological charge 7. Oscillations lead to decay of persistent current as in the case of single-charged systems, but its behavior is more spontaneous and unpredictable.

Three frequencies characterize our system: the rotational quantum, the critical frequency for inner and outer surface modes excitation, and the angular frequency of sound propagating around the ring. The rotational quantum is defined via parameters of the system: $\Omega_0/2\cdot\pi=h/MR^2\approx 1$ Hz. The critical frequency can be obtained from the analysis of small azimuthal perturbations around the stationary state \cite{dubessy_liennard_pedri_perrin_2012}, and it gives (in dimensionless units) $\Omega_c/2\cdot\pi \approx \sqrt{2} \omega_r \mu^{1/6} / (R+\Delta R /2) \approx 17$ Hz, where  $\Delta R$ is a radial width of the ring in dimensionless units. For over-critical frequencies we observe massive vortex nucleation (see Fig. \ref{FIG:evolution3d} (a)). 
The speed of sound in a uniform superfluid fluid at $T \rightarrow 0$ is $c=\sqrt{gn/M}$, where $n$ is a peak density. From here we obtain the angular frequency of sound propagating around the ring $\Omega_s/2\cdot\pi \approx 35$ Hz. Our model works well until the frequency of perturbation is noticeably below the angular frequency of sound; however, when the frequency exceeds it, we have shock waves and vortex turbulence.

The phase-slip edge has clear peak around critical frequency $\Omega_c$ - narrow frequency domain in approximately 5 Hz where persistent current decay under the influence of perturbations with smaller amplitudes (see Fig.\ref{FIG:phaseslip3d} (b)). In Fig. \ref{FIG:bigring} shown phase-slip edge for two different larger ring with smaller critical frequency. $(N_{(a)},N_{(b)}) =(6,2)\cdot 10^5$ atoms in the  harmonic trap with frequencies $(\omega_{r(a)},\omega_{z(a)},\omega_{r(b)},\omega_{z(b)})=2\pi\cdot(123,600,60,2400)$ Hz, and peak-density ring radius $(R_{(a)},R_{(b)})=(47.12,52.04)$ ${\mu}$m; two-particle interaction parameter $g=0.018$,  and dissipation parameter $\gamma=0.0025$, critical frequency estimation that comes from surface model $(\Omega_{c(a)}, \Omega_{c(b)}) \approx  2\cdot\pi (8.5, 6.1)$ Hz. In both cases there is a noticeable or at least slightly noticeable peak near the critical frequency.

Our simulations uncovered some problems in this configuration. The main problems are wave reflection and interference. The time of reaction on perturbation is quite long (a few seconds). The size of the inner circle is big enough to contain many surface modes, and it is the cause of nucleation of a huge number of vortices and antivortices at over-critical frequencies of the perturbing potential. In general, the results are not sufficient enough, they only provide us the edge of persistent current stability area but the final state of evolution beyond it is unpredictable. 

To minimize all these factors, we consider a system with the same topology but in geometry more similar to the experiment \cite{2017NatPh..13..833T}. More on this in the next section.

\section{Wave propagation in elongated 2D/3D condensate}
\label{section:2d}
\begin{figure}[tbp]
	\includegraphics[width = \columnwidth]{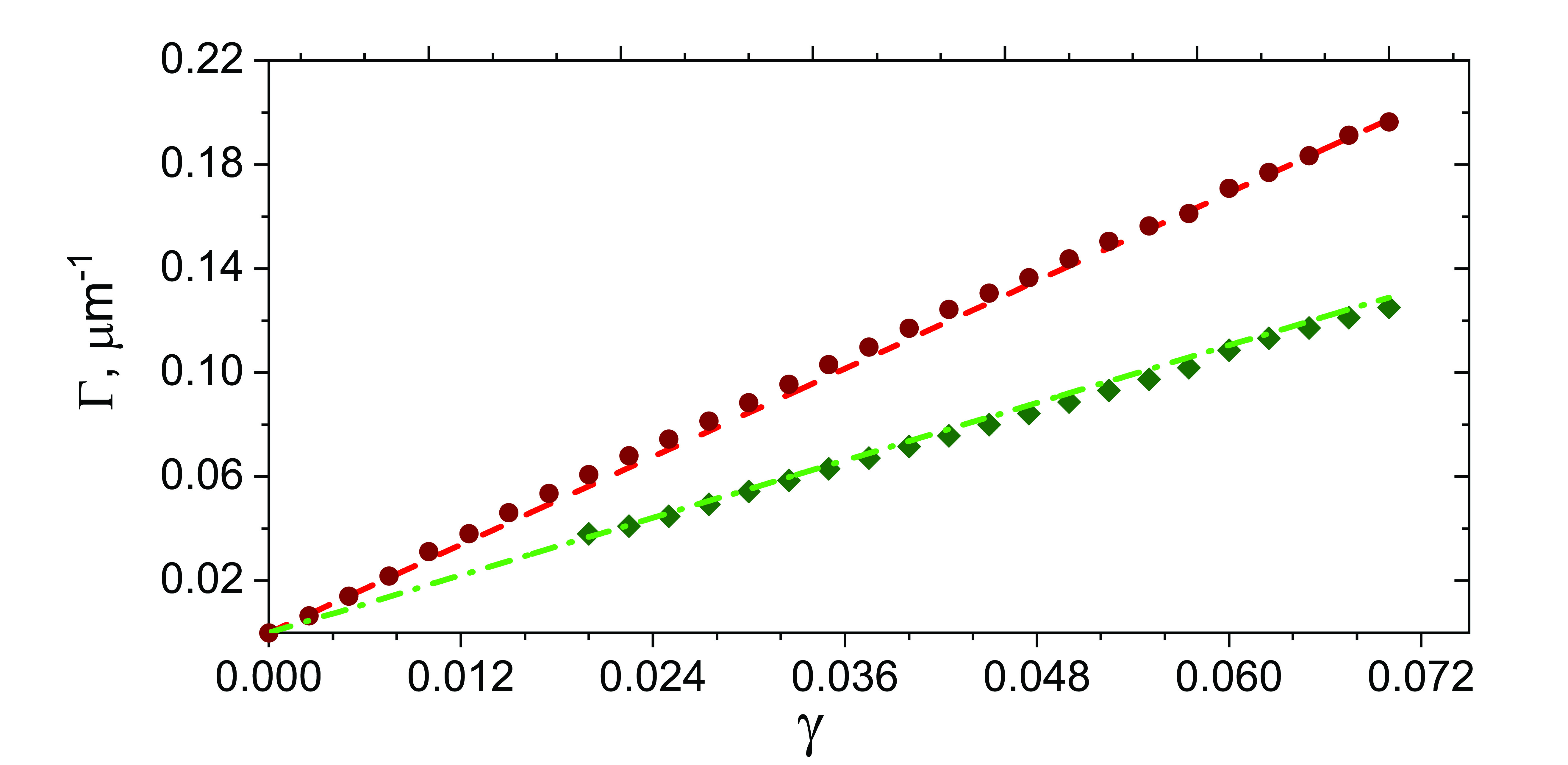}
	\caption{Damping decrement $\Gamma$ for acoustic waves in two(three)-dimensional elongated BEC as a function of the dissipation parameter $\gamma$. Red circles show results of 2D numerical simulations, dashed red line corresponds to the function $\Gamma_{2D} = \gamma/2\xi$ where $\xi_{2D} = 0.177 \mu m$ is healing length in 2D system. Green squares show results of 3D numerical simulations, dash-dot green line corresponds to the function $\Gamma_{3D} = \sqrt{2}\gamma/4\xi$, $\xi_{3D} = 0.192 \mu m$ - healing length in 3D system.}
	\label{FIG:gammaofgamma}
\end{figure}

\begin{figure}[tbp]
	\includegraphics[width = \columnwidth]{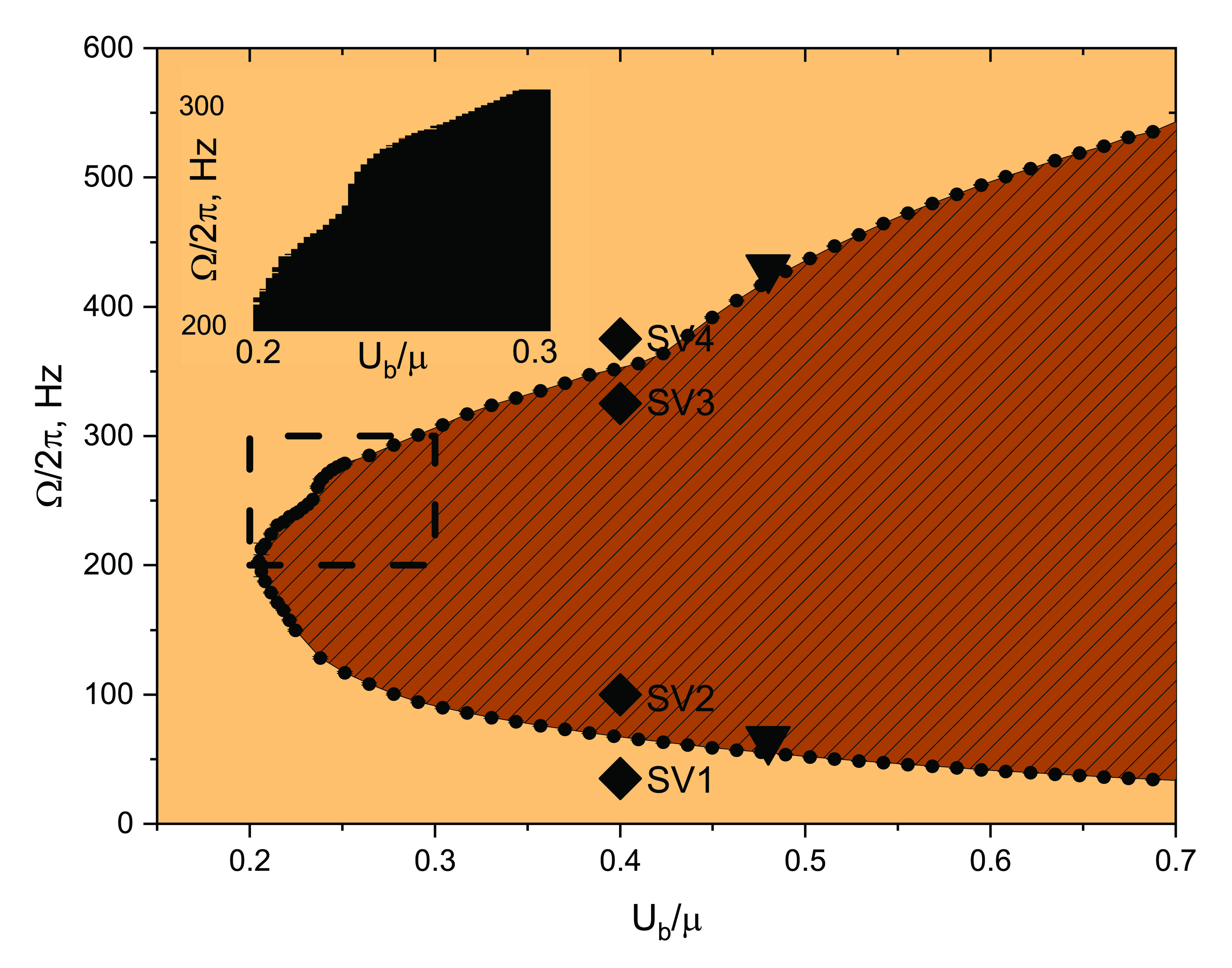}
	\caption{The edge of phase-slip area in space of acoustic wave's parameters (frequency $\Omega$ and amplitude $U_b$). Phase-slip happens (vortex leaves the mini-trap under the influence of acoustic wave) in dashed brown area of $\{\Omega,U_b\}$ parameters. Inset shows the detailed map of final topological charge of mini-trap from small area of parameters bordered with black dashed line. The key stages of evolution with perturbing parameters that correspond to black triangles are shown in  Fig. \ref{FIG:dynamics}. Full simulations for parameters SV1, SV2, SV3 and SV4 can be found in supplemental material videos. }
	\label{FIG:phaseslip}
\end{figure}

In this section we study analogy of water tank: we have elongated along the $x$-axis tiny two-dimensional condensate and pinned in the mini-trap on the $x$-axis vortex thread (See Fig. \ref{FIG:scheme} (c)). Here we can avoid a limitation of the annulus and neglect the interference. In case when the transverse frequency of harmonic trap is high compared to the radial ones $\omega_{z}\gg(\omega_{x},\omega_{y})$ the condensate has a plane-shaped form and can be described by quasi-two-dimensional (2D) wave function $\Psi(x,y,t)$. We assume that the elongated system is tightly confined in a $z$ direction. Therefore, the transverse motion of condensate is frozen, and the wave function in Eq. \ref{EQN:GPE} can be separated into two parts:

	\begin{equation} 
	\label{EQN:separation}
	\psi(\mathbf{r},t)= \Psi(x,y,t)\Xi(z,t),
	\end{equation}
	
	\noindent  where $\Xi(z,t)=(\pi l_{z})^{-1/2}\exp\left(-i\omega_{z}t/2 -z^2/2 l_{z}^2\right)$ is the ground state wave function in the oscillatory  potential $V_{z}(z) = M\omega_z^2z^2/2.$ Here $l_{z}=\sqrt{\hbar/(M\omega_{z})}$ is an oscillatory length in the $z$ direction.

In order to ignore interference effects, we first need to configure the system the way not to let reflected waves interact with a vortex thread. For this purpose we consider two-dimensional elongated along the $x$-axis BEC of $N=1.8 \cdot 10^5$ atoms trapped in harmonic potential in a transverse direction with $\omega_{y}= 2 \pi \cdot 600 $ Hz and in a box-like potential in a longitudinal direction:

	\begin{equation}
	\label{EQN:ext_pot_cig_2nd}
	\begin{aligned}
	V_{trap}(\mathbf{r},t)  = & \frac{M\omega_y^2 y^2}{2} \\ & +  
	U_0 \left[2-\tanh{(x)}+\tanh{(x-L)}\right],
    \end{aligned}
	\end{equation}

	\begin{figure*}[tbp]
	\includegraphics[width = \textwidth]{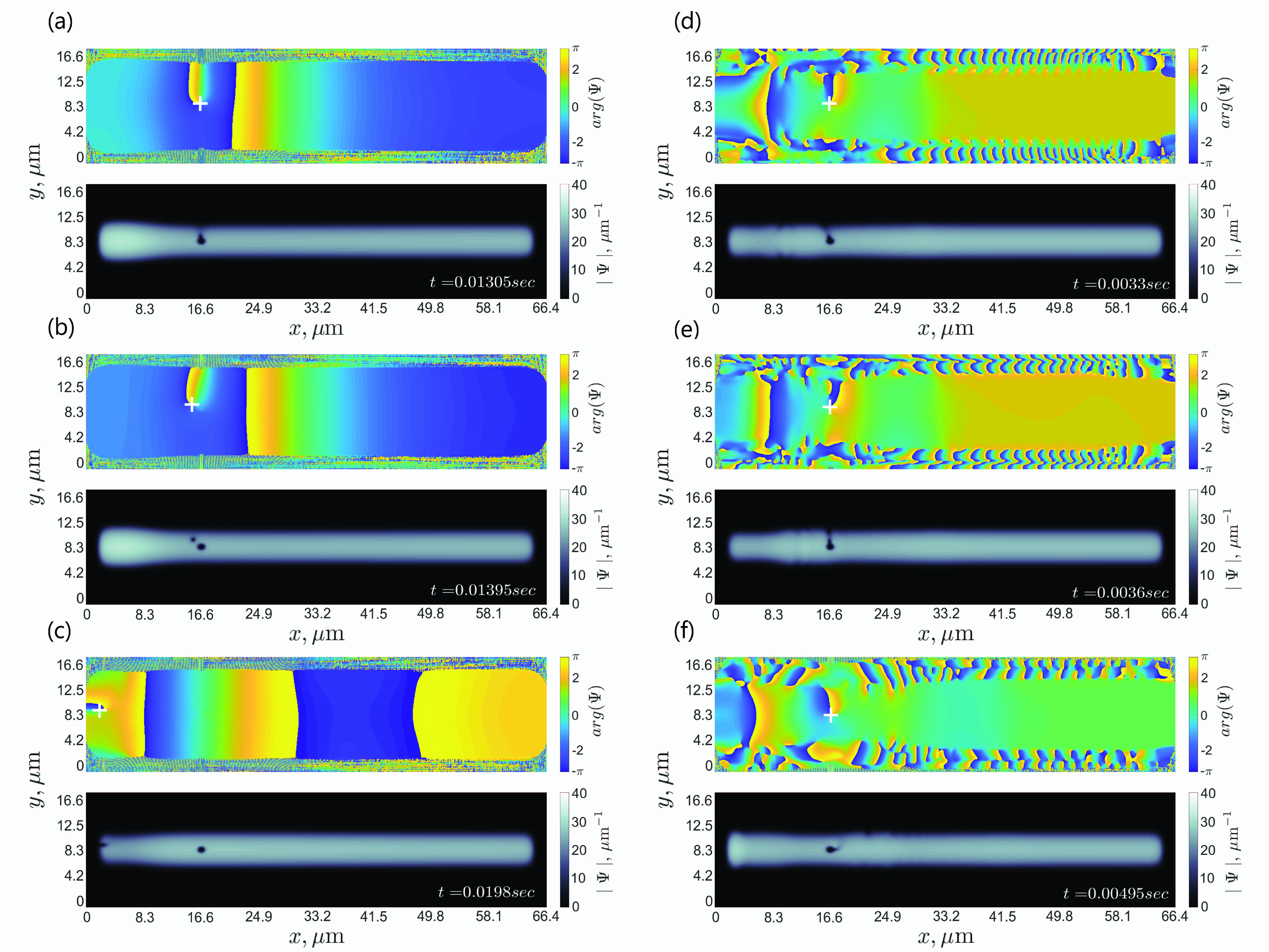}
	\caption{Snapshots (a)-(f) show density and phase distribution evolution of the 2D elongated condensate with periodic perturbation. Both series of shots correspond to the same amplitude of perturbing potential ($U_b/\mu=0.48$) but different frequencies that are shown as black triangles in phase-slip diagram (see Fig. \ref{FIG:phaseslip}). White crosses in phase profiles show the cores of the vortex threads. Time is shown in the right bottom corner of each density profile. (a)-(c) evolution in area \textit{over bottom-edge and under top-edge $\Omega=2\pi\cdot 56$ } Hz: Vortex emits from the mini-trap and converts into outer surface modes; (d)-(f) evolution in area \textit{over top-edge $\Omega=2\pi\cdot 417$ } Hz: Vortex-acoustic wave interaction induce outer surface excitation but vortex emission doesn't occur even in turbulent regime of perturbing potential.}
	\label{FIG:dynamics}
\end{figure*}
	
\begin{figure*}[tbp]
	\includegraphics[width = \textwidth]{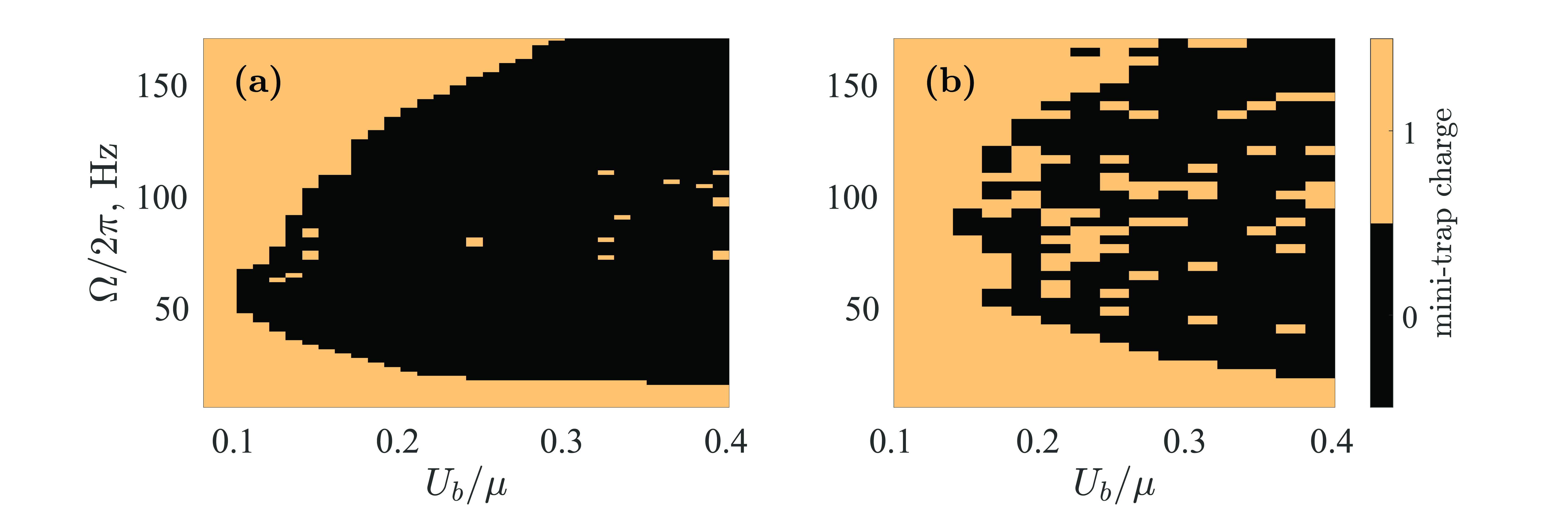}
	\caption{Topological charge of mini-trap in 3D elongated condensate after 0.055 sec evolution without turning off the perturbation. Systems with two different configurations of transverse trap frequencies and number of atoms $(a)$ $N=2.9 \cdot 10^5$,  $(\omega_{y},\omega_{z})= 2 \pi \cdot(123, 600) $Hz and $(b)$ $N=8.3 \cdot 10^5$,  $(\omega_{y},\omega_{z})= 2 \pi \cdot(123, 425) $Hz 
	}
	\label{FIG:cig3dedge}
\end{figure*}

where $L$ is the length of the condensate on x direction is comparable with a size of the chamber from the previous section $L \approx 60 \mu$m, and $U_0$ is the depth of the box-like potential on x-direction. As a source of acoustic waves we use periodic in time Gauss-like perturbation localized around the left edge of the box-like potential:	
	\begin{equation}
	\label{EQN:perturbation_2d}
	V_{pert} (\mathbf{r}) =  U_b  \exp\left\{-\frac{(x-x_0)^2}{\Delta^2} \right\}  \sin (\Omega t),
	\end{equation}
where $U_b$ is the amplitude of the perturbation, $x_0 $ is coordinate of the center of the potential, $\Delta$ - the width of the potential, and $\Omega$ - frequency of the perturbation.
Due to the dissipation, acoustic waves are damping with distance as $e^{-\Gamma x}$. To find the relation between the dissipation parameter $\gamma$ and logarithmic decrement $\Gamma$, we numerically solve the Eq. \ref{EQN:GPE} for single-period excitation. Also, dissipation causes a loss of particles during evolution, so the chemical potential $\mu(t)$ is adjusted at each time step such that the number of condensed particles remains $N$. For a set of different $\gamma$ we save coordinate of density maximum each moment assuming that is the center of the acoustic wave. From these data, we can calculate decrement $\Gamma$ and build $\gamma$ of $\Gamma$ relation in Fig. \ref{FIG:gammaofgamma} as black solid circles.

Keeping the length of the condensate constant, we choose the dissipation parameter $\gamma=0.0175$. For studying acoustic wave - vortex interaction, it is necessary to pin the vortex thread; otherwise, it will not remain in an unstable region of peak density. Therefore, to the trapping potential, we add additional Gauss-like mini-trap potential for keeping imprinted vortices inside of it:   

	\begin{equation}
	\label{EQN:ext_pot_cig_2nd_mt}
	V_{mt}(\mathbf{r},t)  =  U_{mt} \exp\left\{{-\frac{(x-l_{mt})^2+y^2)}{R_{mt}^2}}\right\},
	\end{equation}
where $U_{mt}$ is a depth of the mini-trap ($U_{mt} \gtrsim \mu$), $l_{mt}$ is the distance from the edge of the condensate to the center of the mini-trap, $R_{mt}$ is a radius of the mini-trap for imprinted vortex. The size of the trap is very small $\approx 1 \mu $ m in diameter. The main benefit of such a small size is that the only one surface mode can be excited instead of the torus case when the inner ring surface is a massive vortex factory. 

In this configuration, acoustic wave - vortex interaction happens much faster and more clear. It always takes a few dozens of $\mu$s, and the final state is well-defined. If there is enough energy for exciting inner circle surface mode and emit the vortex, it happens when the first front of the acoustic wave reaches the vortex thread. After leaving the mini-trap, vortex moves to the left(right) edge of the condensate (to the source of perturbation), and when it gets to the low-density area, converts into outer surface mode (see Fig. \ref{FIG:dynamics} (a-c)).

Unlike the previous section where we were calculating angular momentum per atom for finding the edge of persistent current stability, here we detect vortex cores in a small area with a center in the mini trap. The edge of vortex emission from mini-trap is shown in Fig. \ref{FIG:phaseslip}.

The area of amplitude and frequency of the perturbation, which can cause vortex emission, is bounded both by the top and bottom perturbation frequency, as is shown in Fig. \ref{FIG:phaseslip}. For frequencies above the top edge of phase slips the excess energy goes to the creation of a dipole, which almost instantly annihilates with the emission of a sound wave, which can be seen in Fig.~\ref{FIG:dynamics} (f); therefore, this secondary acoustic wave may amplify some modes of initial acoustic wave.

Same has been done in elongated 3D system when z-direction is not frozen (in harmonic trap on $y$ and $z$ directions and in box-like on $x$ direction). Area of perturbation parameters that lead to vortex emission is shown in Fig. \ref{FIG:cig3dedge}. In low frequency/amplitude regime system demonstrates good match with 2D case. More cylinder-like high-density system $(b)$ for high frequencies and amplitudes demonstrates shock wave behavior and vortex turbulence similar to \cite{mossman_hoefer_julien_kevrekidis_engels_2018}. 

\section{Conclusions}
\label{section:conc}

We have investigated persistent current decay driven by acoustic waves in trapped ultracold gases with ring topology. The phase slips are studied in 3D  toroidal trap, quasi-2D and 3D elongated condensate with  vortex lines pinned in a density dips. In all considered cases, acoustic waves were created by the amplitude-modulated repulsive beam. 

Firstly, in the 3D case, it has been found out that the phase-slips can be induced by the acoustic perturbations; however, the wave reflection and interference influence the system essentially. Therefore, even though there is a visible edge of persistent current stability, well above this threshold the final states of the condensate  become practically unpredictable. In toroidal geometry, the system reacts slowly to excitation and due to the presence of many vortices dynamics are complicated. Dissipation eliminates small-scaled noise, but it cannot solve the problem of interfering with a reflected wave: weak dissipation does not remove even short-wavelength excitations, strong dissipation quenches the wave´s amplitude on scales much smaller than the size of the inner ring and, therefore, cannot lead to decay of persistent current.   

On the other hand, the elongated quasi-2D system shows the reliable edge of the phase-slip area in space of acoustic wave parameters (frequency and amplitude). An interesting feature has been observed: the area of phase-slip existence is bounded by the top and bottom wave frequency for the same wave amplitude and this effect might find an application in atomtronics for developing residual-current devices in BEC by analogy with electrical devices but for frequency instead of voltage. It is worth mentioning that the system remains stable for high out-of-edge frequencies.  When side perturbation switches to turbulent regime, it excites outer-surface modes and vortex turbulence. The imprinted vortex thread interacts with them creating dipoles that immediately annihilate with sound emission (that might amplify initial wave) but itself remains locked in mini-trap. The vortex thread can also be emitted with turning into outer surface mode later. Unlike the classic water tank experiment \cite{2017NatPh..13..833T} a quantum vortex in BEC has a small energy. Thus, it may require the use multicharged persistent currents for experimental observation of amplification of acoustic waves in ultracold atomic gases. We hope the results described here can provide a further research direction for studies of quantum properties of rotational superradiance in BEC.



\end{document}